# DietTopp: A first implementation and evaluation of a simplified bandwidth measurement method

Andreas Johnsson, Bob Melander, and Mats Björkman
{Andreas.Johnsson, Bob.Melander, Mats.Bjorkman}@mdh.se
The Department of Computer Science and Engineering
Mälardalen University, Sweden

*Abstract*— This paper describes the active available bandwidth measurement tool DietTopp. It measures the available bandwidth and the link capacity on an end-to-end path having one bottleneck link. DietTopp is based on a simplified TOPP method. This paper describe and motivate the simplifications and assumptions made to TOPP. Further, the paper describes some of the DietTopp implementation issues.

A first evaluation of DietTopp in a testbed scenario is made. Within this evaluation the performance and measurement accuracy of DietTopp is compared to the state-of-the-art tools Pathload and Pathrate.

We show that DietTopp gives fast and accurate estimations of both the available bandwidth and the link capacity of the bottleneck link.

## I. INTRODUCTION

Measurements in best-effort networks are important for network error diagnosis and performance tuning but also as a part of the adaptive machinery of user applications such as streaming video. Within our research we have focused on actively measuring available bandwidth between two network end-points. Such active measurements are done by injecting probe packets (with a pre-defined separation) into the network. The probe packets are time stamped at the receiver end. These time stamps are then used to form an estimate of the available bandwidth. This is discussed in more detail in Section II.

State-of-the-art bandwidth measurement tools and methods are for example TOPP [1], Pathload [2], Pathchirp [3], Delphi [4] and Spruce [5]. An overview of methods and tools in this area can be found in [6].

Within the scope of this paper we have developed, implemented and evaluated a new available bandwidth measurement tool called DietTopp. This tool relies on a simplification of the TOPP bandwidth measurement method [1]. We show that DietTopp gives fast and accurate estimates of both the available bandwidth and the measured link capacity when there is one congested link in the end-to-end path.

This paper is organized as follows. The TOPP measurement method is briefly described in Section II. The simplifications and assumptions made by DietTopp are discussed and motivated in Section III. DietTopp implementation issues are described in Section IV. Section V describes the testbed that has been used to evaluate our DietTopp implementation, while Section VI shows and discusses the obtained results. Section VI also compare DietTopp to other state-of-the-art measurement tools.

## II. TOPP: THE ORIGINAL METHOD

This section briefly describes the original TOPP measurement method that estimates the available bandwidth and link capacity on an end-to-end path. More information on definitions and theory can be obtained from [1].

The TOPP measurement method is divided into two phases, the probing phase and the analysis phase. These two phases are separately described in the following two subsections.

### A. Probe phase

Starting at some offered probe rate $o_{min}$, the TOPP method injects $m$ probe packet trains, where each train contains $k$ equally sized probe packets, into the network path. When all probe trains corresponding to a probe rate $o_{min}$ have been received on the other end, TOPP increases the offered rate $o = o_{min} + \Delta o$. Another set of probe packet trains are sent through the network at the new probe rate. This is repeated until the offered probe rate reaches some specified probe rate $o_{max}$ (i.e. $o > o_{max}$).

The probe packet separation changes between the probe sender and the probe receiver. This is due to the *bottleneck spacing effect* [7] which basically says that the time separation increases in a predictable manner when a link is congested.

The receiver time stamps each packet arrival. Hence, the change in probe packet separation can be measured. The time stamps are then used to calculate the measured probe rate $m$. When the measured probe rate and its corresponding offered probe rate is known the analysis phase of TOPP can be executed. The analysis is described in the next subsection.

*B. Analysis phase*

The analysis builds on comparing the offered probe rate $o_i$ to the measured probe rate $m_i$ on each probe rate level $i$. If plotting the ratio $o_i/m_i$ on the y-axis and $o_i$ on the x-axis for all $i$, we get a plot like the theoretical one in Figure 1. When the network is underloaded the $o_i/m_i$ ratio will be close to $y = 1$. When TOPP increases the offered probe rate some link on the network path will eventually get saturated. Hence, the measured probe rate will decrease since the probe packet separation increases (due to the bottleneck spacing effect). This causes the curve to rise. Segment $b_1$ is linear and the slope corresponds to the link bandwidth of the first congested link. The available bandwidth of the end-to-end path is defined as the intersection of $y = 1$ and the linear segment $b_1$ ($t_1$).

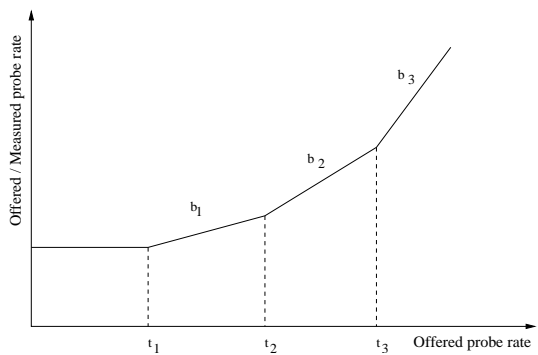

Fig. 1. Plot of offered prob rate / measured probe rate as a function of the offered probe rate. The three segments $b_1$ - $b_3$ corresponds to three congested links while $t_1$ - $t_3$ are breakpoints between congested links.

If there are more than one congested link on the end-to-end path the curve will be divided into several linear segments, when increasing the offered probe rate. Each new segment, $b_2$ and $b_3$, corresponds to the fact that an additional link has been congested. That is, the number of linear segments depends on the number of saturated links on the end-to-end path. The saturation point for each link can be calculated using $t_1$, $t_2$ and $t_3$ in the figure.

The original TOPP method can, in many cases, determine not only the end-to-end available bandwidth (corresponding to the intersection of $y = 1$ and the linear segment $b_1$) but also the link available bandwidth of each congested link. This is done by extracting additional information from the linear segments shown in Figure 1. Exactly how this is done is described in [1].

*C. TOPP complications*

The problem with the original TOPP method is its complex iterative algorithm it uses to estimate the segment intersection points. This makes the TOPP analysis phase computationally expensive. It is feasible, as shown in [1] but takes a lot of computation power.

### III. DIETTOPP

This section describes DietTopp. That is, the simplifications we have made to the TOPP analysis.

DietTopp assumes that only one link between the sender and the receiver is congested. That is, there will only be one linear segment in Figure 1 (i.e. $b_1$). Hence, the end-to-end link capacity is proportional to the slope of segment $b_1$ and the end-to-end available bandwidth is defined as the intersection of $b_1$ and $y = 1$ ($t_1$). That is, the iterative part of TOPP can be omitted.

Since we want to identify $b_1$, the first step is to ensure that DietTopp sends probe packets at a rate above the available bandwidth (corresponding to $t_1$ in Figure 1). This is done by calculating the proportional share, $ps_{max}$, of the end-to-end path at probe rate $o_{max}$. Hence, $ps_{max} = o_{max}/(x + o_{max}) * l$ where $o_{max}$ is the maximum offered probe rate, $x$ the cross traffic and $l$ the link capacity of the congested link. The $ps_{max}$ is estimated by injecting a set of probe packets back-to-back into the network and then measure their separation at the receiver. The $ps_{max}$ is an overestimation of the available bandwidth, as discussed in [1] (and referred to as the asymptotic dispersion rate in [8]).

When DietTopp has obtained a value of $ps_{max}$ it continues the probing phase by injecting probe packets at rates in the interval $[ps_{max}, z * ps_{max}]$, where $z$ is a constant. This will generate $o_i/m_i$ values corresponding to the linear segment $b_1$ in Figure 1. That is, DietTopp can calculate the link capacity by finding the slope of $b_1$ and the available bandwidth by calculating the intersection of $b_1$ and $y = 1$. This is done using linear regression.

We argue that the assumption that only one link is congested in a path is not too far fetched. Usually the bottlenecks are found in the access networks, close to the user, for example when the user is using a wireless connection. Usually wireless links provides a much lower bandwidth than the rest of the links in the path.

Other probing tools that assume one single bottleneck link are Spruce [5] and Delphi [4]. In addition, these tools require prior knowledge of the bottleneck link capacity. This is not the case when using DietTopp. On the contrary, DietTopp will estimate that capacity as part of its estimation procedure.

## IV. IMPLEMENTATION OF DIETTOPP

This section gives an overview of the DietTopp implementation. More information about the implementation ca be obtained by downloading the tool along with its source code [9].

DietTopp is designed for Unix system and is implemented in C++. It has a sender and a receiver part. The sender *probes* the network path by injecting a set of packet trains at different rates similarly to TOPP (in the original method TOPP used pairs of probe packets instead of trains of probe packets). The receiver records the time for each probe packet arrival. These values are sent back to the probe sender for analysis. The analysis is done using our simplified TOPP method described in Sections II and III.

The probe packets used by DietTopp are UDP/IP packets with a size of 1500 bytes. The packets are divided in trains where each train consists of 16 packets. On each probe rate level DietTopp sends 5 trains. DietTopp uses 15 prob rate levels in the interval $[ps_{max}, z * ps_{max}]$ by default.

When DietTopp is measuring the proportional share (described in Section III), 15 trains with 48 packets in each are sent at the maximum probe rate (i.e. the sender's link speed).

The DietTopp implementation can be summarized as follows:

1) Send a set of probe trains at maximum rate
2) Record the probe packet arrival times at receiver
3) Send time stamps back to sender
4) The proportional share ($ps$) is estimated using the arrival times
5) The sender initiates and transmits a set of probe trains with rates in the interval $[1 * ps, 1.5 * ps]$
6) The receiver records the probe packet arrival times
7) The arrival times are sent back to the sender
8) The arrival times are analyzed using our simplified TOPP method
9) If the correlation is large enough and if the standard deviation small enough DietTopp presents its results and terminates
10) Else, repeat from step 1, but increase the number of probe packets in each train

## V. THE TESTBED

The testbed we have used in this work consists of 7 ordinary PC machines. The PCs acting as routers (R1 - R3) are connected by 10 Mbps links while the rest of the links can communicate at 100 Mbps. The testbed is shown in Figure 2. The probe tool sender is running on *S* while the receiver is running on the destination machine *D*.

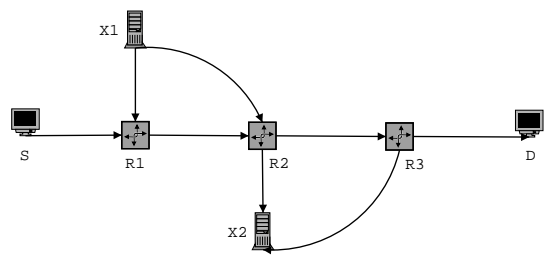

Fig. 2. The testbed used for a first evaluation of DietTopp.

The cross traffic can either take the route $X1 \rightarrow R1 \rightarrow R2 \rightarrow X2$ or the route $X1 \rightarrow R2 \rightarrow R3 \rightarrow X2$. The cross traffic itself is exponentially distributed and consists of 60, 148, 500 and 1500 byte packets (this corresponds to the packet size at the *link layer*). This distribution of packet sizes originates from findings in [10]. The cross traffic is generated by *tg* [11].

An important observation is that the ethernet cards used in the testbed can not send packets back-to-back. The cards add a gap corresponding to 25 bytes between two successive packets. This will be manifested as if all packets will have an increased effective packet size. That is, a 500 byte packet will have an effective packet size of 525 byte when the ethernet card continuously has to forward packets.

## VI. DIETTOPP EVALUATION

This section presents a first evaluation of DietTopp. We have performed measurements in the testbed described in Section V. The cross traffic flows through the route $X1 \rightarrow R1 \rightarrow R2 \rightarrow X2$. That is, the probe packets injected by DietTopp are only affected by cross traffic on one hop (i.e. between one pair or routers). The cross traffic composition and distribution is described in Section V.

We have used the state-of-the-art tools Pathload and Pathrate [2], [12] to compare the accuracy and performance of the measured available bandwidth and link capacity. We show that DietTopp gives accurate and fast estimates. The results are presented and discussed in the following subsection.

*A. Measurement results*

The measurements in Figure 3 originate from DietTopp measurements under four different cross traffic intensities - 0, 3.75, 6.26 and 8.76 Mbps (shown on the x-axis). The y-axis shows the measured link capacity (thick solid line) and measured available bandwidth (thin solid line). The link capacity has a decreasing trend when increasing the cross traffic intensity. Exactly why this happens is subject of further research.

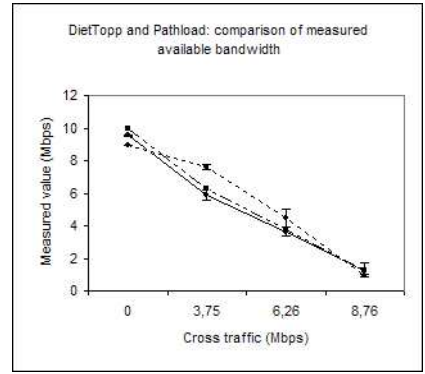

Fig. 5. Available bandwidth measured by DietTopp (solid line) and Pathload (dashed line). The theoretical available bandwidth is shown as the double dot dashed line.

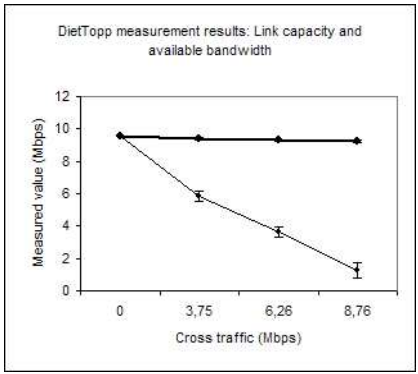

Fig. 3. Link capacity (thick solid line) and available bandwidth (thin solid line) measured by DietTopp.

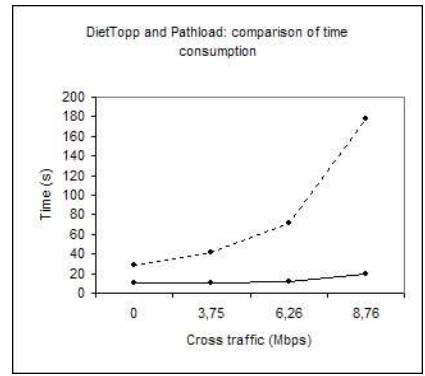

Fig. 6. Time consumption at different cross traffic rates. The solid line corresponds to DietTopp while the dashed line corresponds to Pathload.

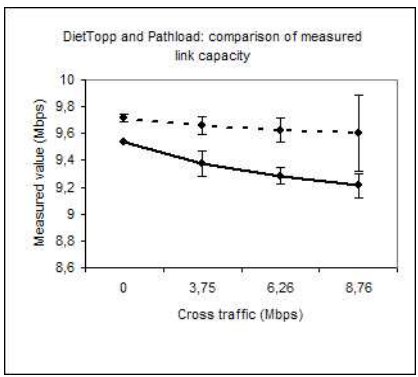

Fig. 4. Link capacity measured by DietTopp (solid line) and Pathrate (dashed line).

The diagram in Figure 4 compares the measured link capacity when using DietTopp (solid line) and Pathrate (dashed line). It is clear that DietTopp underestimates the link capacity in comparison to Pathrates estimation. However, Pathrates estimation is in turn an underestimation compared to the theoretical link capacity of 10 Mbps.

The diagram in Figure 5 is a comparison of the measured available bandwidth. The solid line corresponds to DietTopp while the dashed line corresponds to Pathload. The double dot dashed line is the theoretical available bandwidth under different cross traffic rates. Here we see that both methods estimate the available bandwidth quite well, even if Pathload tends to overestimate in some cases.

The diagram in Figure 6 is a performance comparison of DietTopp and Pathload. The diagram compares the time consumption of DietTopp (solid line) and Pathload (dashed line). As can be seen, the time to measure the available bandwidth using DietTopp is almost constant while for Pathload the time grows exponentially.

It should also be noted that DietTopp does not only measure the available bandwidth during its execution time, but also the link capacity of the congested link.

The diagrams in Figure 7 and 8 compare the number of bits transferred totally during a measurement session and the number of bits transferred per second when using DietTopp (solid line) and Pathload (dashed line). Pathload uses more bits, but since DietTopp runs over a shorter period of time DietTopp is more aggressive as can be seen in the diagram in Figure 8.

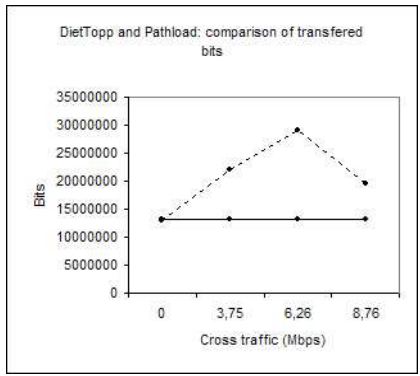

Fig. 7. Number of transfered bits by DietTopp (solid line) and Pathload (dashed line).

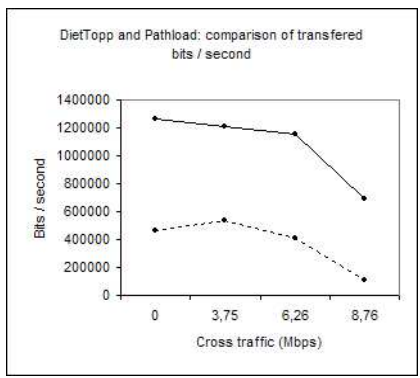

Fig. 8. Number of transferred bits / second by DietTopp (solid line) and Pathload (dashed line).

To summarize, our first evaluation of DietTopp shows that DietTopp estimates both the link capacity and the available bandwidth with comparable and in some cases better accuracy than Pathload and Pathrate. Further, our tool runs over a shorter period of time. The drawback is that DietTopp is more aggressive than both Pathload and Pathrate.

In our continued work we will investigate how the aggressiveness of DietTopp affects TCP flows and other communication protocols. We will also try to make estimations of the available bandwidth and the link capacity with equal accuracy but with fewer probe packets.

## VII. CONCLUSIONS

We have simplified the TOPP measurement method and implemented the new method in a tool that we call DietTopp. We have described and motivated the simplifications and assumptions made to TOPP. Further, we have done an initial evaluation of the accuracy and performance of DietTopp. We have also compared Diet-Topp to the measurement tools Pathload and Pathrate.

We have shown that our tool gives accurate estimate, in the one hop case, of the available bandwidth and an acceptable estimate of the link capacity. Compared to Pathload our solution is quicker, but with the drawback of a higher network aggressiveness.

We will continue our research by investigate how DietTopp reacts when cross traffic is present on multiple links. We will try to find a way to keep the estimation accuracy but decrease the number of probe packets sent. Finally, we will investigate how the accuracy and speed is affected by wireless networks.